\documentclass[prb,twocolumn,groupedaddress,floatfix]{revtex4}
\usepackage{amssymb}
\usepackage{amsmath}
\usepackage{graphicx}
\usepackage{bm}

\newcommand{\nf}{n_{\rm F}}

\newcommand{\jk}{J}

\newcommand{\curH}{{\cal H}}
\newcommand{\tc}{{T_{\rm c}}}
\newcommand{\tk}{T_{\rm K}}

\newcommand{\bk}{{\bf k}}

\newcommand{\be}{\begin{equation}}
\newcommand{\ee}{\end{equation}}
\newcommand{\bea}{\begin{eqnarray}}
\newcommand{\eea}{\end{eqnarray}}
\newcommand{\bse}{\begin{subequations}}
\newcommand{\ese}{\end{subequations}}

\input{epsf}
 
\begin{document}

\title{Kondo-lattice screening in a $d$-wave superconductor}
\author{Daniel E. Sheehy$^{1,2}$ and J\"org Schmalian$^{2}$}
\affiliation{$^{1}$ Department of Physics and Astronomy, Louisiana State University,
Baton Rouge, LA 70803 \\
$^2$ Department of Physics and Astronomy, Iowa State University and Ames
Laboratory, Ames IA 50011}
\date{\today }

\begin{abstract}
We show that local moment screening in a Kondo lattice with $d$-wave
superconducting conduction electrons is qualitatively different from the
corresponding single Kondo impurity case. Despite the conduction-electron
pseudogap, Kondo-lattice screening is stable if the gap amplitude obeys $%
\Delta <\sqrt{T_{\mathrm{K}} D}$, in contrast to the single impurity
condition $\Delta <T_{\mathrm{K}}$ (where $T_{\mathrm{K}}$ is the Kondo
temperature for $\Delta = 0$ and $D$ is the bandwidth). Our theory explains
the heavy electron behavior in the $d$-wave superconductor Nd$_{2-x}$Ce$_{x}$%
CuO$_{4}$.
\end{abstract}

\maketitle





\section{Introduction}

\label{intro}

The physical properties of heavy-fermion metals are commonly attributed to
the Kondo effect, which causes the hybridization of local 4-$f$ and 5-$f$
electrons with itinerant conduction electrons. The Kondo effect for a single
magnetic ion in a metallic host is well understood~\cite{Hewson}. In
contrast, the physics of the Kondo lattice, with one magnetic ion per
crystallographic unit cell, is among the most challenging problems in
correlated electron systems. At the heart of this problem is the need for a
deeper understanding of the stability of collective Kondo screening.
Examples are the stability with respect to competing ordered states
(relevant in the context of quantum criticality~\cite{Coleman01}) or low
conduction electron concentration (as discussed in the so-called exhaustion
problem~\cite{Nozieres98}). In these cases, Kondo screening of the lattice
is believed to be more fragile in comparison to the single-impurity
case. In this
paper, we analyze the Kondo lattice in a host with a $d$-wave conduction
electron pseudogap~\cite{pseudogap}. We demonstrate that Kondo lattice screening is then
significantly more robust than single impurity screening. The unexpected
stabilization of the state with screened moments is a consequence of the
coherency of the hybridized heavy Fermi liquid, i.e. it is a unique lattice
effect. We believe that our results are of relevance for the observed large
low temperature heat capacity and susceptibility of \textrm{Nd}$_{2-x}%
\mathrm{Ce}_{x}\mathrm{CuO}_{4}$, an electron-doped cuprate superconductor%
\cite{Brugger93}.

The stability of single-impurity Kondo screening has been investigated by
modifying the properties of the conduction electrons. Most notably,
beginning with the work of Withoff and Fradkin (WF)~\cite{Withoff90}, the
suppression of the single-impurity Kondo effect by the presence of $d$-wave
superconducting order has been studied. A variety of analytic and numeric
tools have been used to investigate the single impurity Kondo screening in a
system with conduction electron density of states (DOS) $\rho \left( \omega
\right) \propto \left\vert \omega \right\vert ^{r}$, with variable exponent 
$r$ (see Refs.~\onlinecite{Withoff90,Borkowski92,Ingersent96,Ingersent98,Fritz04,Fritz,Vojtareview}). 
Here, $r=1$ corresponds to the case of a $d$-wave superconductor, i.e. is the
impurity version of the problem discussed in this paper. For $r\ll 1$ the
perturbative renormalization group of the ordinary~\cite{Anderson} Kondo problem ($r=0$),
can be generalized\cite{Withoff90}. While the Kondo coupling $J$ is marginal,
a fixed point value $J_{\ast}=r/\rho _{0}$ emerges for finite but small $r$.
Here, $\rho _{0}$ is the DOS for $\omega =D$ with bandwidth $D$.  
Kondo screening only occurs for $J_{\ast}$ and the transition from the
unscreened doublet state to a screened singlet ground state is characterized
by critical fluctuations in time.

\begin{figure}[tbp]
\epsfxsize=9.5cm \vskip.9cm \centerline{\epsfbox{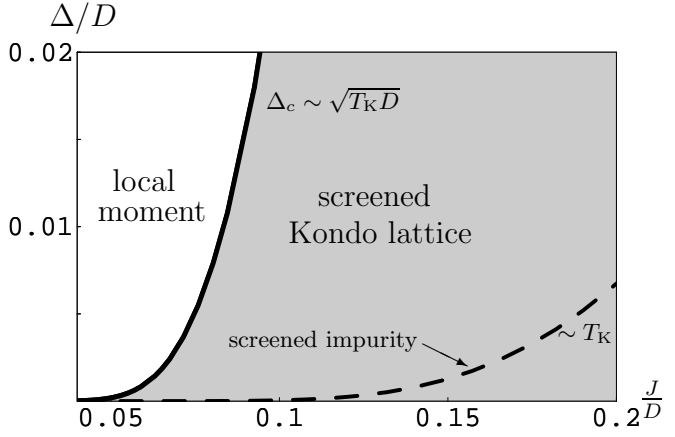}} \vskip-1cm 
\caption{The solid line is the critical pairing strength $\Delta_{c}$ for 
$T\rightarrow 0$ [Eq.~(\protect\ref{eq:deltacone})] separating the Kondo
screened (shaded) and local moment regimes in the Kondo-lattice model Eq.~(%
\protect\ref{eq:ham}). Following well-known results~\protect\cite{Withoff90,Borkowski92}
(see also Appendix~\ref{singleimpuritycase}), the \textit{single-impurity\/} Kondo effect is only
stable for $\Delta \alt D\exp(-2D/J)\sim T_{\mathrm{K}}$ (dashed). }
\label{phasediagram1}
\end{figure}

 Numerical renormalization group (NRG)
calculations demonstrated the existence of a such an impurity quantum
critical point even if $r$ is not small but also revealed that the
perturbative renormalization group breaks down, failing to correctly 
describe this critical
point~\cite{Ingersent98}. For $r=1$, Vojta and Fritz demonstrated that the
universal properties of the critical point can be understood using an
infinite-$U$ Anderson model where the level crossing of the doublet and
singlet ground states is modified by a marginally irrelevant hybridization
between those states\cite{Fritz04,Fritz}. NRG calculations further
demonstrate that the non-universal value for the Kondo coupling at the
critical point is still given by \ $J_{\ast}\simeq r/\rho _{0}$, even if $r$
is not small\cite{Ingersent96}. This result applies to the case of broken
particle-hole symmetry, relevant for our comparison with the Kondo lattice.
In the case of perfect particle hole symmetry it holds that~\cite{Ingersent96} $J_{\ast}\rightarrow
\infty $ for $r\geq 1/2$.

The result  $J_{\ast}\simeq
r/\rho _{0}$ may also be obtained from a large $N$ mean field theory\cite{Withoff90}, 
which otherwise fails to properly describe the critical behavior of the
transition, in particular  if $r$ is not small. The result for $J_{\ast}$
as the transition between the screened and unscreened states relies on the assumption
that the DOS behaves as $\rho \left( \omega \right) \propto
\left\vert \omega \right\vert ^{r}$ all the way to the bandwidth. However, in a
superconductor with nodes we expect that $\rho \left( \omega \right)
\simeq \rho _{0}$ is essentially constant for $|\omega |>\Delta $, with gap
amplitude $\Delta $, altering the predicted location of the transition between
the screened and unscreened states.  To see this, we note that, 
for energies above $\Delta $, the approximately constant DOS implies
the RG flow will be governed by the standard metallic Kondo result~\cite{Anderson,Hewson} with $r=0$, renormalizing
the Kondo coupling to  $\widetilde{J}=J/\left( 1-J\rho _{0}\ln D/\Delta \right) $
with the effective bandwidth $\Delta$ (see Ref.~\onlinecite{Ingersent98}).  Then, we can use the above result
in the renormalized system, obtaining that Kondo screening occurs for  
$\widetilde{J}\rho _{0}\gtrsim r$ 
which is easily shown to be equivalent to the condition $\Delta \lesssim \Delta_*$ with
\be
\Delta_* = e^{1/r}T_{\mathrm{K}},
\label{eq:deltastar}
\ee
where 
\be
\label{eq:tkintro}
T_{\mathrm{K}}=D\exp \left( -\frac{1}{J\rho _{0}}\right), 
\ee
is the Kondo temperature of the system in the absence of pseudogap (which we are
using here to clarify the typical energy scale for $\Delta_*$).  Setting $r=1$ to
establish the implication of Eq.~(\ref{eq:deltastar}) for a $d$-wave
superconductor, we see that, due to the $d$-wave pseudogap in the density of states, the
conduction electrons can only screen the impurity moment if their gap
amplitude is smaller than a critical value of order the corresponding Kondo
temperature $T_{\mathrm{K}}$ for constant density of states. In particular, for $\Delta$
large compared to the (often rather small) energy scale $T_{\mathrm{K}}$, the local moment is 
unscreened, demonstrating the sensitivity
of the single impurity Kondo effect with respect to the low energy behavior
of the host.

Given the complexity of the behavior for a single impurity in a conduction
electron host with pseudogap, it seems hopeless to study the Kondo lattice.
We will show below that this must not be the case and that, moreover, Kondo
screening is stable far beyond the single-impurity result Eq.~(\ref{eq:deltastar}),
as illustrated in Fig.~\ref{phasediagram1} (the dashed line in 
this plot is Eq.~(\ref{eq:deltastar}) with $\rho_0 = 1/2D$).
To do this, we utilize a the large-$N$ mean
field theory of the Kondo lattice to demonstrate that the transition between
the screened and unscreened case is discontinuous. Thus, at least within
this approach, no critical fluctuations occur (in contrast to the single-impurity
case discussed above).
 More importantly, our large-$N$ analysis also finds
 that the stability regime of the Kondo screened lattice is much
larger than that of the single impurity. Thus, the screened heavy-electron state is 
\emph{more} robust and the local-moment phase only emerges if the
conduction electron $d$-wave gap amplitude obeys 
\begin{equation}
\label{eq:deltacapprox}
\Delta >\Delta _{c}\simeq \sqrt{T_{\mathrm{K}}D}\gg T_{\mathrm{K}},
\end{equation}
with $D$ the conduction electron bandwidth. Below, we shall derive a more detailed
expression for $\Delta_c$; in Eq.~(\ref{eq:deltacapprox}) we are simply emphasizing
that $\Delta_c$ is large compared to $T_{\mathrm{K}}$ [and, hence, Eq.~(\ref{eq:deltastar})].

In addition, we find that for $\Delta <\Delta _{c}$, the renormalized mass only 
weakly depends on $\Delta$,  except for the region close to $\Delta _{c}$.  
We give a detailed
explanation for this enhanced stability of Kondo lattice screening,
demonstrating that it is a direct result of the opening of a hybridization
gap in the heavy Fermi liquid state. Since the result was obtained using a
large-$N$ mean field theory we stress that such an approach is not expected to 
 properly describe the detailed nature close to the transition. It should,
however, give a correct order of magnitude result for the location of the
transition.

\begin{figure}[tbp]
\epsfxsize=9.5cm \vskip.9cm \centerline{\epsfbox{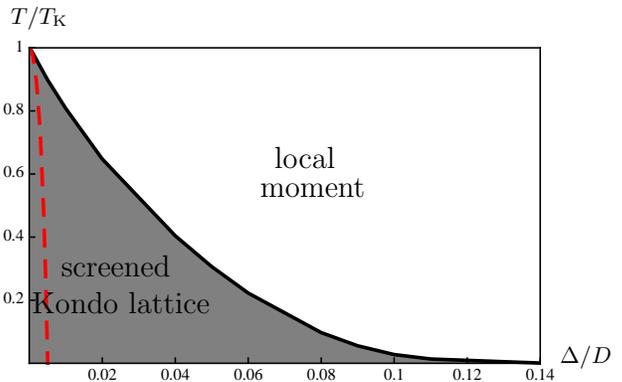}} \vskip-1cm 
\caption{
(Color online) The solid line is a plot of the Kondo temperature $%
T_{\mathrm{K}}(\Delta)$, above which $V = 0$ (and Kondo screening is
destroyed), normalized to its value at $\Delta=0$ [Eq.~(\protect\ref%
{eq:kondotemperature})], as a function of the $d$-wave pairing amplitude $%
\Delta$, for the case of $J = 0.3 D$ and $\protect\mu = -0.1 D$. With these
parameters, $T_{\mathrm{K}}(0) = 0.0014 D$, and $\Delta_c$, the point where $%
T_{\mathrm{K}}(\Delta)$ reaches zero, is $0.14 D$ [given by Eq.~(\protect\ref%
{eq:deltacone})] The dashed line indicates a spinodal, along which the term
proportional to $V^2$ in the free energy vanishes. At very small $\Delta <
2.7\times 10^{-4} D$, where the transition is continuous, the dashed line
coincides with the solid line. }
\label{fig:tvd}
\end{figure}

To understand the resilience of Kondo-lattice screening, recall that, in the
absence of $d$-wave pairing, it is well known that the lattice  Kondo effect (and
concomitant heavy-fermion behavior) is due a hybridization of the conduction
band with an $f$-fermion band that represents excitations of the lattice of
spins. A hybridized Fermi liquid emerges from this interaction. We shall see that, due to the
coherency of the Fermi liquid state, the resulting hybridized heavy fermions
are only {\it marginally\/} affected by the onset of conduction-electron pairing.
This weak proximity effect, with a small $d$-wave gap amplitude $\Delta
_{f}\simeq \Delta T_{\mathrm{K}}/D$ for the heavy fermions, allows the Kondo
effect in a lattice system to proceed via $f$-electron-dominated
heavy-fermion states that screen the local moments, with such screening
persisting up to much larger values of the $d$-wave pairing amplitude than
implied by the single impurity result\cite{Withoff90,Borkowski92}, as
depicted in Fig.~\ref{phasediagram1} (which applies at low $T$). A typical
finite-$T$ phase diagram is shown in Fig.~\ref{fig:tvd}.

Our theory directly applies to the electron-doped cuprate \textrm{Nd}$_{2-x}%
\mathrm{Ce}_{x}\mathrm{CuO}_{4}$, possessing both $d$-wave superconductivity%
\cite{Tsuei00,Prozorov00} with ${T_{\mathrm{c}}}\simeq 20K$ and heavy
fermion behavior below~\cite{Brugger93} $T_{\mathrm{K}}\sim 2-3K$. The
latter is exhibited in a large linear heat capacity coefficient $\gamma
\simeq 4\mathrm{J}/(\mathrm{mol\times K}^{2})$ together with a large
low-frequency susceptibility $\chi $ with Wilson ratio $R\simeq 1.6$. The
lowest crystal field state of \textrm{Nd}$^{3+}$ is a Kramers doublet, well
separated from higher crystal field levels~\cite{Hien98}, supporting Kondo
lattice behavior of the \textrm{Nd}-spins. The superconducting \textrm{Cu-O}%
-states play the role of the conduction electrons. Previous theoretical work
on \textrm{Nd}$_{2-x}\mathrm{Ce}_{x}\mathrm{CuO}_{4}$ discussed the role of
conduction electron correlations\cite{Fulde93}. Careful investigations show
that the single ion Kondo temperature slightly increases in systems with
electronic correlations\cite{Khaliullin95,Hofstetter00}, an effect
essentially caused by the increase in the electronic density of states of
the conduction electrons. However, the fact that these conduction electrons
are gapped has not been considered, even though the Kondo temperature is
significantly smaller than the $d$-wave gap amplitude $\Delta \simeq 3.7%
\mathrm{meV}$ (See Ref.~\onlinecite{Huang90}). We argue that Kondo screening
in \textrm{Nd}$_{2-x}\mathrm{Ce}_{x}\mathrm{CuO}_{4}$ with $T_{\mathrm{K}%
}\ll \Delta $ can only be understood in terms of the mechanism discussed
here.

 We add for completeness that an alternative scenario for the large low
temperature heat capacity of \textrm{Nd}$_{2-x}\mathrm{Ce}_{x}\mathrm{CuO}%
_{4}$ is based on very low lying spin wave excitations\cite{Henggeler98}.
While such a scenario cannot account for a finite value of $C\left( T\right)
/T$ as $T\rightarrow 0$, it is consistent with the shift in the overall
position of the \textrm{Nd}-crystal field states upon doping. However, an
analysis of the spin wave contribution of the \textrm{Nd}-spins shows that
for realistic parameters $C\left( T\right) /T$ vanishes rapidly below the
Schottky anomaly\cite{Bala98}, in contrast to experiments. Thus we believe
that the large heat capacity and susceptibility\ of \textrm{Nd}$_{2-x}%
\mathrm{Ce}_{x}\mathrm{CuO}_{4}$ at low temperatures originates from Kondo
screening of the \textrm{Nd}-spins. 

Despite its relevance for the $d$-wave
superconductor \textrm{Nd}$_{2-x}\mathrm{Ce}_{x}\mathrm{CuO}_{4}$, \ we
stress that our theory does not apply to heavy electron $d$-wave
superconductors, such as $\mathrm{CeCoIn}_{5}$ (see Ref.~%
\onlinecite{Petrovic}), in which the $d$-wave gap is not a property of the
conduction electron host, but a more subtle aspect of the heavy electron
state itself. The latter gives rise to a heat capacity jump at the
superconducing transition $\Delta C\left( T_{c}\right) $ that is comparable
to $\gamma T_{c}$, while in our theory $\Delta C\left( T_{c}\right) \ll
\gamma T_{c}$ holds.

\section{Model}

The principal aim of this paper is to study the screening of local moments
in a $d$-wave superconductor. Thus, we consider the Kondo lattice Hamiltonian,
possessing local spins ($\mathbf{S}_{i}$) coupled to conduction electrons ($%
c_{\mathbf{k}\alpha }$) that are subject to a pairing interaction: 
\begin{equation}
\mathcal{H}=\sum_{\mathbf{k},\alpha }\xi _{\mathbf{k}}c_{\mathbf{k}\alpha
}^{\dagger }c_{\mathbf{k}\alpha }^{{\phantom{\dagger}}}+\frac{J}{2}%
\sum_{i,\alpha ,\beta }\mathbf{S}_{i}\cdot c_{i\alpha }^{\dagger }\bm{\sigma}%
_{\alpha \beta }^{{\phantom{\dagger}}}c_{i\beta }^{{\phantom{\dagger}}}+U_{%
\mathrm{pair}}.  \label{eq:ham}
\end{equation}%
Here, $J$ is the exchange interaction between conduction electrons and local
spins and $\xi _{\mathbf{k}}=\epsilon _{\mathbf{k}}-\mu $ with $\epsilon _{%
\mathbf{k}}$ the conduction-electron energy and $\mu $ the chemical
potential. The pairing term 
\begin{equation}
U_{\mathrm{pair}}=-\sum_{\mathbf{k},\mathbf{k}^{\prime }}U_{\mathbf{k}%
\mathbf{k}^{\prime }}c_{\mathbf{k}\uparrow }^{\dagger }c_{-\mathbf{k}%
\downarrow }^{\dagger }c_{-\mathbf{k}^{\prime }\downarrow }c_{\mathbf{k}%
^{\prime }\uparrow },
\end{equation}%
is characterized by the attractive interaction between conduction electrons $%
U_{\mathbf{k}\mathbf{k}^{\prime }}$. We shall assume the latter stabilizes $%
d $-wave pairing with a gap $\Delta _{\mathbf{k}}=\Delta \cos 2\theta $ with 
$\theta $ the angle around the conduction-electron Fermi surface.

We are particularly interested in the low-temperature strong-coupling phase
of this model, which can be studied by extending the conduction-electron and
local-moment spin symmetry to $SU(N)$ and focusing on the large-$N$ limit~%
\cite{largeN}. In case of the single Kondo impurity, the large-$N$ approach
is not able to reproduce the critical behavior at the transition from a
screened to an unscreeened state. However, it does correctly determine the
location of the transition, i.e. the non-universal value for the strength of
the Kondo coupling where the transition from screened to unscreened impurity
takes place\cite{Ingersent96}. Since the location of the transition and not
the detailed nature of the transition is the primary focus of this paper, a
mean field theory is still useful.

Although the physical case corresponds to $N=2$, the large-$N$ limit yields
a valid description of the heavy Fermi liquid Kondo-screened phase\cite%
{Shiba}. We thus write the spins in terms of auxiliary $f$ fermions as $%
\mathbf{S}_{i}\cdot \mathbf{\sigma }_{\alpha \beta }\rightarrow f_{i\alpha
}^{\dagger }f_{i\beta }-\delta _{\alpha \beta }/2$, subject to the
constraint 
\begin{equation}
\sum_{\alpha }f_{i\alpha }^{\dagger }f_{i\alpha }=N/2.  \label{eq:constraint}
\end{equation}

To implement the large-$N$ limit, we rescale the exchange coupling via $%
J/2\rightarrow J/N$ and the conduction-electron interaction as $U_{\mathbf{k}%
,\mathbf{k}^{\prime }}\rightarrow s^{-1}U_{\mathbf{k},\mathbf{k}^{\prime }}$
[where $N\equiv (2s+1)$]. The utility of the large-$N$ limit is that the
(mean-field) stationary-phase approximation to $\mathcal{H}$ is believed to
be exact at large $N$. Performing this mean field decoupling of $\mathcal{H}$
yields 
\begin{eqnarray}
&&\hspace{-.5cm}\mathcal{H}=\sum_{\mathbf{k},m=-s}^s\Big[\xi _{\mathbf{k}}c_{%
\mathbf{k}m}^{\dagger }c_{\mathbf{k}m}+V\left( f_{\mathbf{k}m}^{\dagger }c_{%
\mathbf{k}m}+h.c.\right) +\lambda f_{\mathbf{k}m}^{\dagger }f_{\mathbf{k}m}%
\Big]  \notag \\
&&-\sum_{\mathbf{k,}m=1/2}^{s}\left( \Delta _{\mathbf{k}}^{\dagger }c_{-%
\mathbf{k}-m}c_{\mathbf{k}m}+h.c.\right) +E_{0},  \label{(eq:Hmf)}
\end{eqnarray}
with $E_{0}$ a constant in the energy that is defined below. The pairing
gap, $\Delta _{\mathbf{k}}$, and the hybridization between conduction and $f$%
-electrons, $V$, result from the mean field decoupling of the pairing and
Kondo interactions, respectively. The hybridization $V$ (that we took to be
real) measures the degree of Kondo screening (and can be directly measured
experimentally~\cite{Optical}) and $\lambda $ is the Lagrange multiplier
that implements the above constraint, playing the role of the $f$-electron
level. The free energy $F$ of this single-particle problem can now be
calculated, and has the form: 
\begin{eqnarray}
&&F(V,\lambda ,\Delta _{\mathbf{k}})=\frac{NV^{2}}{J}-\frac{N\lambda }{2}
+s\sum_{\mathbf{k}\mathbf{k}^{\prime }}\Delta _{\mathbf{k}}\Delta _{\mathbf{%
k }^{\prime }}U_{\mathbf{k}\mathbf{k}^{\prime }}^{-1}  \label{eq:ffullygen}
\\
&&+N\sum_{\mathbf{k,}\alpha \mathbf{=\pm }}\left( \frac{1}{4}(\xi
_{k}+\lambda )-\frac{1}{2}E_{\mathbf{k}\alpha }-T\ln \left( 1+\mathrm{e}%
^{-\beta E_{\mathbf{k}\alpha }}\right) \right) ,  \notag
\end{eqnarray}
where $T=\beta ^{-1}$ is the temperature. The first three terms are the
explicit expressions for $E_{0}$ in Eq.~(\ref{(eq:Hmf)}), and $E_{\mathbf{k}%
\pm }$ is 
\begin{eqnarray}
E_{\mathbf{k}\pm } &=&\frac{1}{\sqrt{2}}\sqrt{\Delta _{\mathbf{k}%
}^{2}+\lambda ^{2}+2V^{2}+\xi _{\mathbf{k}}^{2}\pm \sqrt{S_{\mathbf{k}}}},
\label{eq:epm} \\
S_{\mathbf{k}} &=&(\Delta _{\mathbf{k}}^{2}+\xi _{\mathbf{k}}^{2}-\lambda
^{2})^{2}+4V^{2}\left[ (\xi _{\mathbf{k}}+\lambda )^{2}+\Delta _{\mathbf{k}%
}^{2}\right] ,  \notag
\end{eqnarray}%
describing the bands of our $d$-wave paired heavy-fermion system. 

The phase
behavior of this Kondo lattice system for given values of $T$, $J$ and $\mu $
is determined by finding points at which $F$ is stationary with respect to
the variational parameters $V$, $\lambda $, and $\Delta _{\mathbf{k}}$. For
simplicity, henceforth we take $\Delta _{\mathbf{k}}$ as given (and having $%
d $-wave symmetry as noted above) with the goal of studying the effect of
nonzero pairing on the formation of the heavy-fermion metal characterized by 
$V$ and $\lambda$ that satisfy the stationarity conditions 
\begin{subequations}
\label{Eq:stat}
\begin{eqnarray}
\frac{\partial F}{\partial V} = 0, \\
\frac{\partial F}{\partial \lambda} = 0,  \label{eq:constraintdiff}
\end{eqnarray}
with the second equation enforcing the constraint, Eq.~(\ref{eq:constraint}%
). We shall furthermore restrict attention to $\mu <0$ (i.e., a less than
half-filled conduction band).

\begin{figure}[tbp]
\epsfxsize=9.5cm \vskip.9cm \centerline{\epsfbox{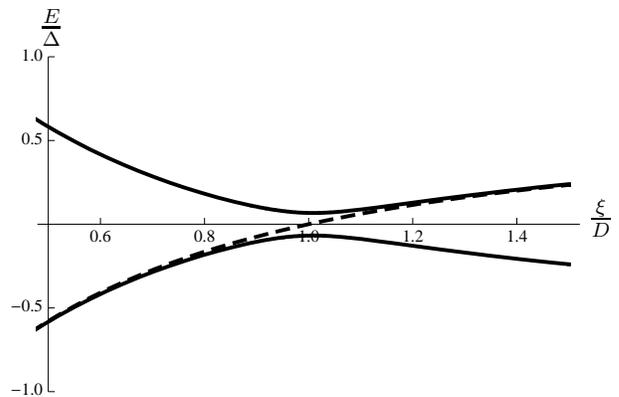}} \vskip-1cm 
\caption{The dashed line is the lower heavy-fermion band (crossing zero at the
heavy-fermion Fermi surface) for the unpaired ($\Delta = 0$) case 
and the solid lines are $\pm E_{\bk -}$ for $\Delta_\bk = 0.1D$, showing
a small f-electron gap $\Delta_{f{\bk}} \simeq .014D$. }
\label{fig:gapplot}
\end{figure}

Before we proceed we point out that the magnitude of the pairing gap near
the unpaired heavy-fermion Fermi surface (located at $\xi =V^{2}/\lambda $)
is remarkably small. Taylor expanding $E_{\mathbf{k}-}$ near this point, we
find 
\end{subequations}
\begin{equation}
E_{\mathbf{k}-}\simeq \frac{\lambda ^{2}}{V^{2}}\left[ \left( \xi
-V^{2}/\lambda -\lambda \Delta _{\mathbf{k}}^{2}/V^{2}\right) ^{2}+\Delta _{%
\mathbf{k}}^{2}\right] ^{1/2},
\end{equation}
giving a heavy-fermion gap $\Delta _{f\mathbf{k}}=\left( \lambda /V\right)
^{2}\Delta _{\mathbf{k}}$ [with amplitude $\Delta_f = \Delta \left( \lambda
/V\right)^2$]. We show below that $\left( \lambda /V\right) ^{2}$ $\ll 1$
such that $\Delta _{f\mathbf{k}}\ll \Delta _{\mathbf{k}}$. In 
Fig.~\ref{fig:gapplot}, 
we plot the lower heavy-fermion band for the unpaired
case $\Delta _{\mathbf{k}}=0$ (dashed line) along with $\pm E_{\mathbf{k}-}$
for the case of finite $\Delta _{\mathbf{k}}$ (solid lines) in the vicinity
of the unpaired heavy-fermion Fermi surface, showing the small heavy-fermion
gap $\Delta _{f\mathbf{k}}$. Thus, we find a weak proximity effect in which
the heavy-fermion quasiparticles, which are predominantly of $f$-character,
are only weakly affected by the presence of $d$-wave pairing in the
conduction electron band.

\begin{figure}[tbp]
\epsfxsize=9.5cm \vskip.9cm \centerline{\epsfbox{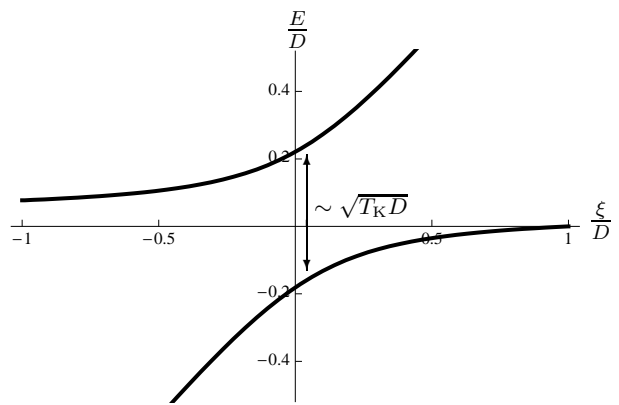}} \vskip-1cm 
\caption{Plot of the energy bands $E_{+}(\xi)$ (top curve)  and  $E_{-}(\xi)$ 
(bottom curve), defined in Eq.~(\ref{eq:eplusminus}), in the heavy Fermi 
liquid state (for $\Delta = 0$), for the case $V = 0.2D$ and $\lambda = 0.04 D$, that has a heavy-fermion Fermi surface near $\xi = D$ 
and an experimentally-measurable hybridization gap~\cite{Optical}
 (the minimum value of $E_{+}-E_{-}$, i.e., the direct gap) equal to $2V\sim \sqrt{\tk D}$.  Note, however, the 
indirect gap is $\lambda \sim \tk$.}
\label{fig:energybands}
\end{figure}

\section{Kondo lattice screening}

\subsection{Normal conduction electrons}

A useful starting point for our analysis is to recall the well-known\cite%
{Coleman} unpaired ($\Delta =0$) limit of our model. 
By minimizing the correpsonding free energy [simply the $\Delta = 0$ limit of Eq.~(\ref{eq:ffullygen})], one obtains,
at low temperatures, that
the Kondo screening of the local moments is represented by the nontrivial
stationary point of $F$ at $V=V_{0}$ and $\lambda =\lambda _{0}=V_{0}^{2}/D$, with 
\begin{equation}
V_{0}\simeq \sqrt{\frac{D+\mu }{2\rho _{0}}}\exp \left( -\frac{1}{2J\rho _{0}%
}\right) ,  \label{eq:rfinalzeroh}
\end{equation}
Here we have taken the conduction electron density of states to be a
constant, $\rho _{0}=(2D)^{-1}$, with $2D$ the bandwidth. The resulting
phase is a metal accommodating both the conduction and $f$-electrons with a
large density of states $\propto \lambda _{0}{}^{-1}$ near the Fermi surface
 at $\epsilon _{\mathbf{k}}\simeq \mu +V_{0}^{2}/\lambda _{0}$, revealing its
heavy-fermion character.  In Fig.~\ref{fig:energybands}, we plot the 
energy bands 
\begin{equation}
E_{\pm }\left( \xi _{\mathbf{k}}\right) =\frac{1}{2}\left( \xi_{\mathbf{k}
}+\lambda \pm \sqrt{\left( \xi _{\mathbf{k}}-\lambda \right) ^{2}+4V^{2}}
\right) ,\label{eq:eplusminus}
\end{equation}
of this heavy Fermi liquid in the low-$T$ limit. 

With increasing $T$, the stationary $V$ and $\lambda$ decrease monotonically, vanishing at the Kondo temperature 
\begin{eqnarray}
T_{\mathrm{K}} &=&\frac{2\mathrm{e}^{\gamma }}{\pi }\sqrt{D^{2}-\mu ^{2}}%
\exp \big[-\frac{1}{\rho _{0}J}\big],  \label{eq:kondotemperature} \\
&=&\frac{2\mathrm{e}^{\gamma }}{\pi }\sqrt{\frac{D-\mu }{D+\mu }}\lambda
_{0}.
\end{eqnarray}%
Here, the second line is meant to emphasize that $T_{\mathrm{K}}$ is of the
same order as the $T=0$ value of the $f$-fermion chemical potential $\lambda_0$, 
and therefore $T_{\mathrm{K}}\ll V_{0}$, i.e., $T_{\mathrm{K}}$ is small
compared to the zero-temperature hybridization energy $V_{0}$.

It is well established that the phase transition-like behavior of $V$ at $T_{%
\mathrm{K}}$ is in fact a crossover once $N$ is finite~\cite{Hewson,largeN}.
Nevertheless, the large-$N$ approach yields the correct order of magnitude
estimate for $T_{\mathrm{K}}$ and provides a very useful description of the
strong coupling heavy-Fermi liquid regime, including the emergence of a
hybridization gap in the energy spectrum.

\subsection{ $d$-wave paired conduction electrons}

Next, we analyze the theory in the presence of $d$-wave pairing with gap
amplitude $\Delta $. Thus, we imagine continuously turning on the $d$-wave
pairing amplitude $\Delta $, and study the stability of the Kondo-screened
heavy-Fermi liquid state characterized by the low-$T$ hybridization $V_{0}$,
Eq.~(\ref{eq:rfinalzeroh}). As we discussed in Sec.~\ref{intro}, in the case
of a \textit{single\/} Kondo impurity, it is well known that Kondo screening
is qualitatively different in the case of $d$-wave pairing, and the single
impurity is only screened by the conduction electrons if the Kondo coupling
exceeds a critical value
\begin{equation}
J_{\ast}\simeq \frac{1}{\rho _{0}}\frac{1}{1+\ln D/\Delta }.  \label{eq:Jc}
\end{equation}
For $J<J_{\ast}$, the impurity is unscreened. This result for $J_{\ast}$ can
equivalently be expressed in terms of a critical pairing strength $\Delta_{\ast}$, 
beyond which Kondo screening is destroyed for a given $J$: 
\begin{equation}
\Delta _{\ast }=D\exp \big[1-\frac{1}{\rho _{0}J}\big],  \label{Eq:deltastar}
\end{equation}
[equivalent to  Eq.~(\ref{eq:deltastar}) for $r=1$],
which is proportional to the Kondo temperature $T_{\mathrm{K}}$. This
result, implying that a $d$-wave superconductor can only screen a local spin
if the pairing strength is much smaller than $T_{\mathrm{K}}$, can also be
derived within the mean-field approach to the Kondo problem, as shown in
Appendix~\ref{singleimpuritycase} (see also Ref.~\onlinecite{Borkowski92}).
Within this approach, a continuous transition to the unscreened phase (where $
V^{2}\rightarrow 0$ continuously) takes place at $\Delta \simeq \Delta_{\ast}$.

\begin{figure}[tbp]
\epsfxsize=9.5cm \vskip.9cm \centerline{\epsfbox{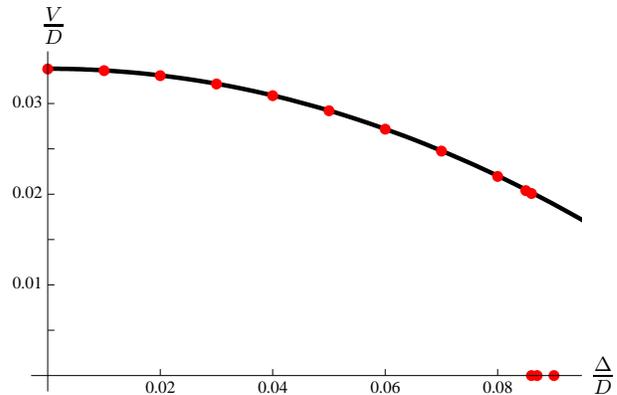}} \vskip-1cm 
\caption{(Color online) Main: Mean-field Kondo parameter $V$ as a function of the d-wave pairing amplitude $\Delta$, 
for exchange coupling $\jk = 0.30 D$ and chemical potential $\mu = -0.1D$, according
to the approximate formula Eq.~(\ref{eq:rnoughtquad}) (solid line) and via a direct minimization
of Eq.~(\ref{eq:ffullygen}) at $T=10^{-4} D$ (points), the latter exhibiting a first-order transition near
$\Delta = 0.086 D$. }
\label{fig:rplot1}
\end{figure}

Thus, calculations for the single impurity case indicate that Kondo
screening is rather sensitive to a $d$-wave pairing gap. The question we wish
to address is, how does $d$-wave pairing affect Kondo screening in the lattice
case? In fact, we will see that the results are quite different in the
Kondo lattice case, such that Kondo screening persists beyond the point $%
\Delta_*$. To show this, we have numerically studied the $\Delta$-dependence
of the saddle point of the free energy Eq.~(\ref{eq:ffullygen}), showing
that, at low temperatures, $V$ only vanishes, in a discontinuous manner, at
much larger values of $\Delta$, as shown in Fig.~\ref{fig:rplot1} (solid
dots) for the case of $J = 0.30D$, $\mu = -0.1 D$ and $T = 10^{-4} D$ (i.e., $%
T/T_{\mathrm{K}} \simeq .069$). In Fig.~\ref{fig:tvd}, we plot the phase
diagram as a function of $T$ and $\Delta$, for the same values of $J$ and $%
\mu$, with the solid line denoting the line of discontinuous transitions.

The dashed line in Fig.~\ref{fig:tvd} denotes the spinodal $T_{\mathrm{s}}$
of the free energy $F$ at which the quadratic coefficient of Eq.~(\ref%
{eq:ffullygen}) crosses zero. The significance of $T_{\mathrm{s}}$ is that,
if the Kondo-to-local moment transition were continuous (as it is for $%
\Delta =0$), this would denote phase boundary; the $T\rightarrow 0$ limit of
this quantity coincides with the single-impurity critical pairing 
Eq.~(\ref{Eq:deltastar}). An explicit formula for $T_{\mathrm{s}}$ can be easily obtained
by finding the quadratic coefficient of Eq.~(\ref{eq:ffullygen}): 
\begin{equation}
\frac{1}{J}=\sum_{\mathbf{k}}\frac{\tanh E_{\mathbf{k}}/2T_{\mathrm{s}%
}(\Delta )}{2E_{\mathbf{k}}},  \label{eq:tkvd}
\end{equation}
with $E_{\mathbf{k}}\equiv \sqrt{\xi _{\mathbf{k}}^{2}+\Delta _{\mathbf{k}%
}^{2}}$, and where we set $\lambda =0$ [which must occur at a continuous
transition where $V\rightarrow 0$, as can be seen by analyzing Eq.~(\ref%
{eq:constraintdiff})]. As seen in Fig.~\ref{fig:tvd}, the spinodal
temperature is generally much smaller than the true transition temperature;
however, for very small $\Delta \rightarrow 0$, $T_{\mathrm{s}}(\Delta )$
coincides with the actual transition (which becomes continuous), as noted in
the figure caption.

Our next task is to understand these results within an approximate analytic
analysis of Eq.~(\ref{eq:ffullygen}); before doing so, we stress again that
the discontinuous transition from a screened to an unscreened state as
function of $T$ becomes a rapid crossover for finite $N$. The large $N$
theory is, however, expected to correctly determine where this crossover
takes place.

\subsubsection{Low-$T$ limit}

According to the numerical data (points) plotted in Fig.~\ref{fig:rplot1},
the hybridization $V$ is smoothly suppressed with increasing pairing
strength $\Delta $ before undergoing a discontinuous jump to $V=0$. To
understand, analytically, the $\Delta $-dependence of $V$ at low-$T$, we shall analyze the
 $T=0$ limit of $F$, i.e., the ground-state energy $E$.  The essential question concerns
 the stability of the Kondo-screened state with respect to a $d$-wave pairing gap, characterized
by the following $\Delta$-dependent hybridization
\begin{equation}
V(\Delta) =V_{0}\Big( 1-\frac{\Delta ^{2}}{\Delta _{\mathrm{typ}}^{2}}\Big) ,
\label{eq:vofdelta}
\end{equation}
with $\Delta_{\mathrm{typ}}$ an energy scale, to be derived, that gives the typical
value of $\Delta$ for which the heavy-fermion state is  affected by $d-$wave pairing.

 To show that Eq.~(\ref{eq:vofdelta}) correctly describes the smooth suppression of the 
hybrization with increasing $\Delta$, and to obtain the scale $\Delta_{\rm typ}$,
we now consider the dimensionless quantity  
\begin{equation}
\label{chidef}
\chi_{\Delta }\equiv -\frac{1}{2\rho _{0}}\frac{\partial ^{2}E}{\partial\Delta ^{2}},
\end{equation}
that characterizes the change of the ground state energy with respect to the
pairing gap. Separating the amplitude of the gap from its momentum
dependence, i.e. writing $\Delta _{\mathbf{k}}=\Delta \phi _{\mathbf{k}}$,
we obtain from the Hellmann-Feynman theorem that:
\begin{eqnarray}
\chi _{\Delta } &=&-\frac{1}{2\rho _{0}\Delta }\left\langle \frac{\partial \curH}{\partial \Delta }\right\rangle,   \notag \\
&=&-\frac{N}{2\rho _{0}\Delta }\sum_{\mathbf{k}}\phi _{\mathbf{k}
}\left\langle c_{\mathbf{k}m}^{\dagger }c_{-\mathbf{k-}m}^{\dagger
}\right\rangle .
\end{eqnarray}
For $\Delta \rightarrow 0$ this yields 
\begin{equation}
\chi _{\Delta }=\frac{N}{2\rho _{0}}\int \frac{d\omega }{2\pi }\sum_{\mathbf{
k}}\phi _{\mathbf{k}}^{2}G_{cc}\left( \mathbf{k,}i\omega \right)
G_{cc}\left( -\mathbf{k,-}i\omega \right) .
\end{equation}
Here, $G_{cc}\left( \mathbf{k,}i\omega \right) $ is the conduction electron
propagator. As expected, $\chi _{\Delta }$ is the particle-particle correlator
of the conduction electrons. \ Thus, for $T=0$ the particle-particle response
will be singular. This is the well known Cooper instability. For $V=0$ we
obtain for example 
\begin{equation}
\chi _{\Delta }\left( V=0\right) =\frac{N}{8}\log \frac{D^{2}-\mu ^{2}}{%
\Delta ^{2}},
\end{equation}%
where we used $\Delta $ as a lower cut off to control the Cooper logarithm.
Below we will see that, except for extremely small values of $\Delta $, the
corresponding Cooper logarithm is overshadowed by another logarithmic term
that does not have its origin in states close to the Fermi surface, but
rather results from states with typical energy $V\simeq \sqrt{T_{\mathrm{K}}D}$.

In order to evaluate $\chi _{\Delta }$ in the heavy Fermi liquid state, we
start from: 
\begin{equation}
G_{cc}\left( \mathbf{k,}\omega \right) =\frac{v_{\mathbf{k}}^{2}}{\omega
-E_{+}\left( \xi _{\mathbf{k}}\right) }+\frac{u_{\mathbf{k}}^{2}}{\omega
-E_{-}\left( \xi _{\mathbf{k}}\right) },
\end{equation}
where $E_{\pm}$ is given in Eq.~(\ref{eq:eplusminus}) and
the coherence factors of the hybridized Fermi liquid are: 
\begin{eqnarray}
u_{\mathbf{k}}^{2} &=&\frac{1}{2}\left( 1-\frac{\xi _{\mathbf{k}}-\lambda }{%
\sqrt{\left( \xi _{\mathbf{k}}-\lambda \right) ^{2}+4V^{2}}}\right) ,  \notag
\\
v_{\mathbf{k}}^{2} &=&\frac{1}{2}\left( 1+\frac{\xi _{\mathbf{k}}-\lambda }{%
\sqrt{\left( \xi _{\mathbf{k}}-\lambda \right) ^{2}+4V^{2}}}\right) .
\end{eqnarray}
Inserting $G_{cc}\left( \mathbf{k,}\omega \right) $ into the above
expression for $\chi _{\Delta }$ yields%
\begin{equation}
\chi _{\Delta }=\frac{N}{8}\int_{-D-\mu }^{D-\mu }d\xi \left( \frac{v^{4}}{%
E_{+}}+\frac{u^{4}}{\left\vert E_{-}\right\vert }+\frac{4v^{2}u^{2}\theta
\left( E_{-}\right) }{E_{+}+E_{-}}\right) .  \label{eq:chiD}
\end{equation}%
We used that $E_{+}>0$ is always fulfilled, as we consider a less than half
filled conduction band.

Considering first the limit $\lambda =0$, it holds $E_{-}\left( \xi \right) <0
$ and the last term in the above integral disappears. The remaining terms
simplify to 
\begin{eqnarray}
\chi _{\Delta }\left( \lambda =0\right)  &=&\frac{N}{8}\int_{-D-\mu }^{D-\mu
}d\xi \frac{1}{\sqrt{\xi ^{2}+4V^{2}}},  \notag \\
&=&\frac{N}{8}\log \frac{D^{2}-\mu ^{2}}{4V^{2}}.  \label{eq:lam=0}
\end{eqnarray}
Even for $\lambda $ nonzero, this is the dominant contribution to $\chi
_{\Delta }$ in the relevant limit $\lambda \ll V\ll D$. To demonstrate this
we analyze Eq.~(\ref{eq:chiD}) for nonzero $\lambda$, but assuming $\lambda \ll
V$ as is indeed the case for small $\Delta $. The calculation is lengthy but
straightforward. It follows:
\begin{equation}
\label{eq:chiwithcooper}
\chi_{\Delta }=\frac{N}{8}\left( 1+\frac{\lambda }{D}\right) \log \frac{%
D^{2}-\mu ^{2}}{4V^{2}}+\frac{N}{8}\frac{\lambda }{D}\log \frac{D\left\vert
\mu \right\vert }{\Delta ^{2}} .
\end{equation}
The last term is the Cooper logarithm, but now in the heavy fermion state.
The prefactor $\lambda /D\simeq T_{K}/D$ is a result of the small weight of
the conduction electrons on the Fermi surface (i.e. where $\xi \simeq
V^{2}/\lambda $) as well as the reduced velocity close to the heavy electron
Fermi surface. Specifically it holds $u^{2}\left( \xi \simeq V^{2}/\lambda
\right) \simeq \lambda ^{2}/V^{2}$ as well as $E_{-}\left( \xi \simeq
V^{2}/\lambda \right) \simeq \frac{\lambda ^{2}}{V^{2}}\left( \xi -\frac{%
V^{2}}{\lambda }\right)$.

Thus, except for extremely small gap values where $\Delta ^{2}<D^{2}\left( 
\frac{D}{4T_{K}}\right) ^{-D/T_{K}}$, $\chi _{\Delta }$ is dominated by the $
\lambda =0$ result, Eq.~(\ref{eq:lam=0}), and the Cooper logarithm plays no role
in our analysis. The logarithm in Eq.~(\ref{eq:lam=0}) is not originating from
the heavy electron Fermi surface (i.e. it is not from $\xi \simeq \frac{r^{2}}{\lambda }$ ). 
Instead, it has its origin in the integration over states
where $E_{-}<0$. The important term $\frac{v^{4}}{2E_{+}}-\frac{u^{4}}{2E_{-}
}$ in Eq.~(\ref{eq:chiD}) is peaked for $\xi \simeq 0$ i.e. where $E_{\pm
}\left( \xi \simeq 0\right) =\pm V$ and is large as long as $\left\vert \xi
\right\vert \lesssim V$. For $\xi \simeq 0$ holds $\frac{v^{4}}{2E_{+}}
\simeq -\frac{u^{4}}{2E_{-}}\simeq \frac{1}{32V}$. This peak at $\xi \simeq
0 $ has its origin in the competition between two effects. Usually, $u$ or $v
$ are large when $E_{\pm }\simeq \xi $. The only regime where $u$ or $v$ are
still sizable while $E_{\pm }$ remain small is close to the bare conduction
electron Fermi surface at $\left\vert \xi \right\vert \simeq V$ (the
position of the level repulsion between the two hybridizing bands). \ Thus,
the logarithm is caused by states that are close to the {\it bare\/} conduction electron
Fermi surface.  Although these states have the strongest response to 
a pairing gap, they don't have much to do with the heavy fermion character of the system.  It is
interesting that this  heavy fermion pairing response is the same even in
case of a Kondo insulator where $\lambda =0$ and the Fermi level is in the
middle of the hybridization gap.

The purpose of the preceding analysis was to derive an accurate expression 
for the ground-state energy  $E$ at small $\Delta$.  
Using Eq.~(\ref{chidef}) gives:
\begin{equation}
E=E(\Delta =0) -\chi _{\Delta }\rho _{0}\Delta ^{2},
\end{equation}
which, using Eq.~(\ref{eq:lam=0}) and considering the leading order in $\lambda \ll V$ and $\Delta \ll V$,
safely neglecting the last term of Eq.~(\ref{eq:chiwithcooper}) according to the argument of the previous
paragraph, and dropping overall constants, yields
\begin{equation}
\frac{E}{N}\simeq \frac{V^{2}}{J}-\frac{\lambda }{2}+V^{2}\rho _{0}\ln \frac{%
\lambda }{D+\mu }-\frac{\rho _{0}\Delta ^{2}}{8}\ln \frac{D^{2}-\mu ^{2}}{%
V^{2}}.  \label{eq:approxgse}
\end{equation}
Using Eq.~(\ref{Eq:stat}), the stationary value of the hybridization (to
leading order in $\Delta ^{2}$) is then obtained via minimization with
respect to $V$ and $\lambda $. This yields 
\begin{equation}
V(\Delta )\simeq V_{0}-\frac{\Delta ^{2}}{16V_{0}},  \label{eq:rnoughtquad}
\end{equation}
with the stationary value of $\lambda =2\rho _{0}V^{2}$, which establishes Eq.~(\ref{eq:vofdelta}). 
A smooth
suppression of the Kondo hybridization from the $\Delta =0$ value $V_{0}$
[Eq.~(\ref{eq:rfinalzeroh})] occurs with increasing $d$-wave pairing
amplitude $\Delta $ at low $T$. This result thus implies that the conduction
electron gap only causes a significant reduction of $V$ and $\lambda $ for $%
\Delta \simeq \Delta _{\mathrm{typ}}\propto \sqrt{T_{\mathrm{K}}D}$.

\begin{figure}[tbp]
\epsfxsize=9.5cm \vskip.9cm \centerline{\epsfbox{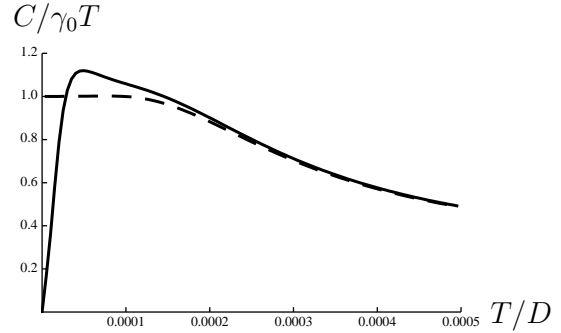}} \vskip-1cm 
\caption{Plot of the low-temperature specific heat coefficient $\frac{C}{T}
= - \frac{\partial^2F}{\partial T^2}$, for the case of $\protect\lambda =
10^{-2} D$, $V = 10^{-1} D$, and $\protect\mu= -0.1 D$, for the metallic
case ($\Delta = 0$, dashed line) and the case of nonzero $d$-wave pairing ($%
\Delta= 0.1 D$, solid line). This shows that, even with nonzero $\Delta$,
the specific heat coefficient will appear to saturate at a large value at
low $T$ (thus exhibiting signatures of a heavy fermion metal), before
vanishing at asymptotically low $T \ll \Delta_f$ ($=\Delta(\protect\lambda%
/V)^2 = 10^{-4} D$) Each curve is normalized to the $T= 0$ value for the
metallic case, $\protect\gamma_0 \simeq \frac{2}{3}\protect\pi^{2}\protect%
\rho_{0}V^{2}/\protect\lambda^{2}$.}
\label{fig:heat}
\end{figure}

In Fig.~\ref{fig:rplot1} we compare $V(\Delta)$ of Eq.~(\ref{eq:rnoughtquad}) 
(solid line) with the numerical result (solid dots). As long as $V$ stays
finite, the simple relation Eq.~(\ref{eq:rnoughtquad}) gives an excellent
description of the heavy electron state. Above the small $f$-electron gap $%
\Delta _{f}$, these values of $V$ and $\lambda $ yield a large heat capacity
coefficient (taking $N=2$) $\gamma \simeq \frac{2}{3}\pi ^{2}\rho
_{0}V^{2}/\lambda ^{2}$ and susceptibility $\chi \simeq 2\rho
_{0}V^{2}/\lambda ^{2}$, reflecting the heavy-fermion character of this
Kondo-lattice system even in the presence of a $d$-wave pairing gap.
According to our theory, this standard heavy-fermion behavior (as observed
experimentally~\cite{Brugger93} in Nd$_{2-x}$Ce$_{x}$CuO$_{4}$) will be
observed for temperatures that are large compared to the $f$-electron gap $%
\Delta_f$. However, for very small $T\ll \Delta _{f}$, the temperature
dependence of the heat capacity changes (due to the $d$-wave character of the $%
f$-fermion gap), behaving as $C=AT^{2}/\Delta $ with a large prefactor $%
A\simeq \left(D/T_{\mathrm{K}}\right)^{2}$. This leads to a sudden drop in
the heat capacity coefficient at low $T$, as depicted in Fig.~\ref{fig:heat}.

The surprising robustness of the Kondo screening with respect to $d$-wave
pairing is rooted in the weak proximity effect of the $f$-levels and the
coherency as caused by the formation of the hybridization gap. Generally, a
pairing gap  affects states with energy $\Delta _{\mathbf{k}}$ from the
Fermi energy. However, low energy states that are within $T_{\mathrm{K}}$ of
the Fermi energy are predominantly of $f$-electron character (a fact that
follows from our large-$N$ theory but also from the much more general Fermi
liquid description of the Kondo lattice~\cite{Yamada}) and are protected by
the weak proximity. These states only sense a gap $\Delta _{f\mathbf{k}}\ll \Delta _{%
\mathbf{k}}$ and can readily participate in local-moment screening. 

Furthermore,
the opening of the hybridization gap coherently pushes conduction electrons
to energies $\simeq V$ from the Fermi energy. Only for $\Delta \simeq V$ $%
\simeq \sqrt{T_{\mathrm{K}}D}$ will the conduction electrons ability to
screen the local moments be affected by $d$-wave pairing. This situation is
very different from the single impurity Kondo problem where conduction
electron states come arbitrarily close to the Fermi energy.

\subsubsection{First-order transition}

The result Eq.~(\ref{eq:rnoughtquad}) of the preceding subsection 
strictly applies for $\Delta \to 0$, although as seen in Fig.~\ref{fig:rplot1},
in practice it agrees quite well with the numerical minimization of 
the free energy until the first-order transition. To understand the way in which $V$ is destroyed with
increasing $\Delta$, we must consider the $V\to 0$ limit of the free energy.

We start with the ground-state energy. Expanding $E$ [the $T \to 0$ limit of Eq.~(\ref{eq:ffullygen})]
to leading order in $V$
and zeroth order in $\lambda $ (valid for $V\rightarrow 0$), we find
(dropping overall constants) 
\begin{equation}
\frac{E}{N}\simeq -4\rho_{0}V^{2}\ln \frac{\Delta_{c}}{\Delta }+\frac{16}{3}
\frac{\rho_{0}}{\Delta }V^{3},  \label{eq:approxenergytrans}
\end{equation}
where we defined the quantity $\Delta_{c}$ 
\begin{equation}
\Delta_{c}=4\sqrt{D^{2}-\mu^{2}}\exp \left( -\frac{1}{2\rho_{0}J}\right),
\label{eq:deltacone}
\end{equation}
at which the minimum value of $V$ in Eq.~(\ref{eq:approxenergytrans})
vanishes continuously,
with the formula for $V(\Delta)$ given by
\be
V(\Delta) \simeq \frac{1}{2} \Delta \ln \frac{\Delta_c}{\Delta},
\label{eq:statsolution}
\ee
near the transition.
According to Eq.~(\ref{eq:deltacone}), the 
equilibrium hybridization $V$ vanishes (along with the destruction of 
Kondo screening)  for pairing
amplitude $\Delta_{c}\sim \sqrt{T_{\mathrm{K}}D}$, of the same order of
magnitude as the $T=0$ hybridization $V_{0}$, as expected [and advertised
above in Eq.~(\ref{eq:deltacapprox})].

Equation~(\ref{eq:deltacone}) strictly applies only at $T=0$, apparently
yielding a continuous transition at which $V\rightarrow 0$ for $\Delta
\rightarrow \Delta _{c}$. What about $T\neq 0$? We find that, for small but
nonzero $T$, Eq.~(\ref{eq:deltacone}) approximately yields the correct
location of the transition, but that the nature of the transition changes
from continuous to first-order. Thus, for $\Delta $ near $\Delta _{c}$,
there is a discontinuous  jump to the local-moment phase that is
best obtained numerically, as shown above in Figs.~\ref{fig:rplot1} and \ref{fig:tvd}. 
However, we can get an approximate analytic understanding of this
first-order transition by examining the low-$T$ limit. Since excitations are
gapped, at low $T$ the free energy $F_{\rm K}$
 of the Kondo-screened ($V\neq 0$) phase
is well-approximated by inserting the stationary solution Eq.~(\ref{eq:statsolution}) 
into Eq.~(\ref{eq:approxenergytrans}): 
\begin{equation}
\frac{F_{\mathrm{K}}}{N}\simeq -\frac{1}{6}\rho _{0}\Delta ^{2}\ln ^{3}\frac{%
\Delta_{c}}{\Delta},  \label{eq:gsenearfake}
\end{equation}
for $F_{\rm K}$ at $\Delta \to \Delta_c$.
The discontinuous Kondo-to-local moment transition occurs when the Kondo
free energy Eq.~(\ref{eq:gsenearfake}) is equal to the local-moment free
energy. For the latter we set $V=\lambda =0$ in Eq.~(\ref{eq:ffullygen}),
obtaining (recall $E_{\mathbf{k}}=\sqrt{\xi _{\mathbf{k}}^{2}
+\Delta _{\mathbf{k}}^{2}}$) 
\begin{eqnarray}
&&\hspace{-0.5cm}\frac{F_{\mathrm{LM}}}{N}\simeq -\frac{1}{2}\rho _{0}(D+\mu
)^{2}-\frac{1}{4}\rho _{0}\Delta ^{2}\ln \frac{4\sqrt{D^{2}-\mu ^{2}}}{%
\Delta }  \notag \\
&&\hspace{-0.5cm}\qquad \quad -T\ln 2-T\sum_{\mathbf{k}}\ln \big[1+\mathrm{e}%
^{-\beta E_{\mathbf{k}}}\big],  \label{eq:localmomentfree}
\end{eqnarray}%
where we dropped an overall constant depending on the conduction-band
interaction.

The term proportional to $T$ in Eq.~(\ref{eq:localmomentfree}) comes from
the fact that $E_{\mathbf{k} -} = 0$ for $V = \lambda = 0$, and corresponds
to the entropy of the local moments. At low $T$, the gapped nature of the
$d$-wave quasiparticles implies the last term in Eq.~(\ref{eq:localmomentfree}%
) can be neglected (although the nodal quasiparticles give a subdominant
power-law contribution). In deriving the Kondo free energy $F_{\mathrm{K}}$,
Eq.~(\ref{eq:gsenearfake}), we dropped overall constant terms; 
re-establishing these to allow a comparison to $F_{\mathrm{LM}}$ , and setting 
$F_{\mathrm{LM}} = F_{\mathrm{K}}$, we find 
\begin{equation}
\frac{1}{6}\rho_0 \Delta^2 \ln^3 \frac{\Delta_{c}}{\Delta} = T\ln 2,
\label{eq:deltajump}
\end{equation}
that can be solved for temperature to find the transition temperature $T_K$
for the first-order Kondo screened-to-local moment phase transition: 
\begin{equation}
\label{eq:tkdeltafirst}
T_K(\Delta) = \frac{\rho_0 \Delta^2}{6\ln 2} \ln^3\frac{\Delta_{c}}{\Delta},
\end{equation}
that is valid for $\Delta \to \Delta_c$, providing an accurate approximation
to the numerically-determined $T_{\mathrm{K}}$ curve in Fig.~\ref{fig:tvd}
(solid line) in the low temperature regime (i.e., near $\Delta_c = 0.14D$ in
Fig.~\ref{fig:tvd}).  

Equation~(\ref{eq:tkdeltafirst}) yields the temperature at which, within mean-field theory,
the screened Kondo lattice is destroyed by the presence of nonzero $d$-wave pairing;
thus, as long as $\Delta< T_K(\Delta)$, heavy-fermion behavior is compatible with 
$d$-wave pairing in our model.  The essential feature of this result is that 
$\tk(\Delta)$ is only marginally reduced from the $\Delta=0$ Kondo temperature
Eq.~(\ref{eq:tkintro}), establishing the stability of this state.  In comparison,
according to expectations based on a single-impurity analysis, one would expect 
the Kondo temperature to follow the dashed line in Fig.~\ref{fig:tvd}.

Away from this approximate result valid at large $N$, the RKKY interaction
between moments is expected to lower the local-moment free energy, altering the predicted
location of the phase boundary. Then,
even for $T=0$, a level crossing between the screened and unscreened ground
states occurs for a finite $V$. Still, as long as the $\Delta =0$ heavy
fermion state is robust, it will remain stable at low $T$ for $\Delta $
small compared to $\Delta_{c}$, as summarized in Figs.~\ref{phasediagram1}
and \ref{fig:tvd}.

\section{Conclusions}

We have shown that a lattice of Kondo spins coupled to an itinerant
conduction band experiences robust Kondo screening even in the presence of $d
$-wave pairing among the conduction electrons. The heavy electron state is
protected by the large hybridization energy $V\gg T_{\mathrm{K}}$. The $d$-wave 
gap in the conduction band induces a relatively weak gap at the
heavy-fermion Fermi surface, allowing Kondo screening and heavy-fermion
behavior to persist. Our results demonstrate the importance of Kondo-lattice
coherency, manifested by the hybridization gap, which is absent in case of dilute
Kondo impurities. As pointed out in detail, the origin for the unexpected
robustness of the screened heavy electron state is the coherency of the
Fermi liquid state. With the opening of a hybridization gap, conduction
electron states are pushed to energies of order $\sqrt{T_{{\normalsize K}}D}$
away from the Fermi energy. Whether or not these conduction electrons open
up a $d$-wave gap is therefore of minor importance for the stability of the
heavy electron state.

Our conclusions are based on a large-$N$ mean field theory. In case of a
single impurity, numerical renormalization group calculations demonstrated
that such a mean field approach fails to reproduce the correct critical
behavior where the transition between screened and unscreened impurity takes
place. However the mean field theory yields the correct value for the
strength of the Kondo coupling at the transition. In our paper we are not
concerned with the detailed nature in the near vicinity of the transition.
Our focus is solely the location of the boundary between the heavy Fermi liquid
and unscreened local moment phase, and we do expect that a mean field theory
gives the correct result. One possibility to test the results of this paper
is a combination of dynamical mean field theory and numerical
renormalization group for the pseudogap Kondo lattice problem.

In case where Kondo screening is inefficient and $\Delta >\sqrt{T_{{\normalsize K}}D}$,
i.e., the \lq\lq local moment\rq\rq\ phase of Figs.~\ref{phasediagram1}  and  \ref{fig:tvd},
 the ground state of the moments will likely be
magnetically ordered. This can have interesting implications for the
superconducting state. Examples are reentrance into a normal phase (similar
to $\mathrm{ErRh}_{4}\mathrm{B}_{4}$, see Ref.~\onlinecite{Fertig77}) or a
modified vortex lattice in the low temperature magnetic phase. In our theory
we ignored these effects. This is no problem as long as the superconducting
gap amplitude $\Delta $ is small compared to $\sqrt{T_{{\normalsize K}}D}$ and
the Kondo lattice is well screened. Thus, the region of stability of the
Kondo screened state will not be significantly affected by including the
magnetic coupling between the $f$-electrons. Only the nature of the
transition and, of course, the physics of the unscreened state will depend
on it.  Finally, our theory offers an explanation for the heavy fermion
state in Nd$_{2-x}$Ce$_{x}$CuO$_{4}$, where $\Delta \gg T_{\mathrm{K}}$.  

\smallskip \noindent \textit{Acknowledgments\/} --- 
We are grateful for useful discussions with A. Rosch and M. Vojta. This
research was supported by the Ames Laboratory, operated for the U.S.
Department of Energy by Iowa State University under Contract No.
DE-AC02-07CH11358. DES was also supported at the KITP under NSF grant
PHY05-51164. 

\appendix  

\section{Single impurity case}

\label{singleimpuritycase}

For a single Kondo impurity a critical value $J_{\ast}$ for the coupling
between conduction electron and impurity spin emerges, separating
Kondo-screened from local moment behavior for a single spin impurity in a
$d$-wave superconductor, see Eq.~(\ref{eq:Jc}). As discussed in the main text,
this is equivalent to a critical pairing Eq.~(\ref{Eq:deltastar}) above
which Kondo screening does not occur. The result was obtained in careful
numerical renormalization group calculations\cite{Ingersent96,Ingersent98}.
In the present section, we demonstrate that the same result also follows
from a simple large-$N$ mean field approach. It is important to stress that
this approach fails to describe the detailed critical behavior. However, here we are
only concerned with the approximate value of the non-universal quantity $J_{\ast}$. 
Indeed, mean field theory is expected to give a reasonable value for
the location of the transition.

Our starting point is the model Hamiltonian 
\begin{eqnarray}
&&\mathcal{H}=\sum_{\mathbf{k}m}\epsilon _{\mathbf{k}}c_{\mathbf{k}%
m}^{\dagger }c_{\mathbf{k}m}^{\phantom{\dagger}}++\frac{J}{N}%
\sum_{m,m^{\prime },\mathbf{k},\mathbf{k}^{\prime }}f_{m}^{\dagger }f_{m}^{%
\phantom{\dagger}}c_{\mathbf{k}m}^{\dagger }c_{\mathbf{k}^{\prime }m^{\prime
}}^{\phantom{\dagger}}  \notag \\
&&\qquad -\sum_{\mathbf{k}\mathbf{k}^{\prime }}U_{\mathbf{k}\mathbf{k}%
^{\prime }}c_{\mathbf{k}\uparrow }^{\dagger }c_{-\mathbf{k}\downarrow
}^{\dagger }c_{-\mathbf{k}^{\prime }\downarrow }^{\phantom{\dagger}}c_{%
\mathbf{k}^{\prime }\uparrow }^{\phantom{\dagger}}.
\end{eqnarray}%
with the corresponding mean-field action $S=S_{f}+S_{b}+S_{\mathrm{int}}$
with (introducing the Lagrange multiplier $\lambda $ and hybridization $V$
as usual, and making the BCS mean-field approximation for the conduction
fermions): 
\begin{eqnarray}
&&\hspace{-0.5cm}S_{f}=\int d\tau \sum_{m}\big[\sum_{\mathbf{k}}c_{\mathbf{k}%
m}^{\dagger }(\partial _{\tau }+\epsilon _{\mathbf{k}})c_{\mathbf{k}m}^{%
\phantom{\dagger}}+f_{m}^{\dagger }(\partial _{\tau }+\lambda )f_{m}^{%
\phantom{\dagger}}\big],  \notag \\
&&\hspace{-0.5cm}S_{b}=\int d\tau \Big(\frac{N}{J}V^{\dagger }V-\lambda
Nq_{0}\Big), \\
&&\hspace{-0.5cm}S_{\mathrm{int}}=\int d\tau \sum_{m\mathbf{k}}\big(%
f_{m}^{\dagger }c_{\mathbf{k}m}^{\phantom{\dagger}}V+V^{\dagger }c_{\mathbf{k%
}m}f_{m}\big)+\sum_{\mathbf{k}\mathbf{k}^{\prime }}\Delta _{\mathbf{k}%
}\Delta _{\mathbf{k}}^{\prime }U_{\mathbf{k}\mathbf{k}^{\prime }}^{-1} 
\notag \\
&&\quad -\sum_{m=1/2}^{J}\big(\Delta _{\mathbf{k}}^{\dagger }c_{-\mathbf{k}%
-m}^{\phantom{\dagger}}c_{\mathbf{k}m}^{\phantom{\dagger}}+c_{\mathbf{k}%
m}^{\dagger }c_{-\mathbf{k}-m}^{\dagger }\Delta _{\mathbf{k}}\big),
\end{eqnarray}%
where the $\lambda $ integral implements the constraint $Nq_{0}=%
\sum_{m}f_{m}^{\dagger }f_{m}$, with $q_{0}=1/2$. Here, we have taken the
large $N$ limit, with $N=2J+1$.

The mean-field approximation having been made, it is now straightforward to
trace over the fermionic degrees of freedom to yield 
\begin{widetext}
\bea
F = \frac{N|V|^2}{\jk} - \lambda N q_0 
- \frac{N}{2} T \sum_\omega 
\ln \Big[ (i\omega - \lambda - \Gamma_1(i\omega)) (i\omega + \lambda + \Gamma_1(-i\omega) ) 
- \Gamma_2(i\omega)  \bar{\Gamma}_2(i\omega)     \Big],
\label{eq:fsingle}
\eea
\end{widetext}
for the free energy contribution due to a single impurity in a $d$-wave
superconductor. Here, we dropped an overall constant due to the conduction
fermions only, as well as the quadratic term in $\Delta_\mathbf{k}$ (which
of course determines the equilibrium value of $\Delta_\mathbf{k}$; here, as
in the main text, we're interested in the impact of a given $\Delta_\mathbf{k}
$ on the degree of Kondo screening), and defined the functions 
\begin{eqnarray}
\Gamma_1(i\omega) &=& |V|^2 \sum_\mathbf{k} \frac{i\omega + \epsilon_\mathbf{%
k}}{(i\omega)^2 -E_{\mathbf{k}}^2}, \\
\Gamma_2(i\omega) &=& V^2 \sum_\mathbf{k} \frac{\Delta_\mathbf{k}}{%
(i\omega)^2 -E_{\mathbf{k}}^2}, \\
\bar{\Gamma}_2(i\omega) &=& (V^\dagger)^2 \sum_\mathbf{k} \frac{\Delta_%
\mathbf{k}}{(i\omega)^2 -E_{\mathbf{k}}^2}.
\end{eqnarray}

At this point we note that, for a $d$-wave superconductor, $\Gamma_2 = \bar{
\Gamma}_2=0$ due to the sign change of the $d$-wave order parameter. The
self-energy $\Gamma_1(i\omega)$ is nonzero and essentially measures the
density of states (DOS)  $\rho_d(\omega)$ of the $d$-wave superconductor. In fact,
one can show that the corresponding retarded function $\Gamma_{1R}(\omega)$
satisfies 
\begin{equation}
\Gamma_{1R}(\omega) = |V|^2 \int_{-\infty}^\infty dz \frac{\rho_d(z)}{\omega
+ i\delta - z} ,
\end{equation}
with $\delta = 0^+$, so that the imaginary part 
$\Gamma_{1R}''(\omega)  = -\pi|V|^2 \rho_d(\omega)$
measures the DOS. Writing $\Gamma_{1R}(\omega) \equiv
|V|^2 G(\omega)$, we have for the free energy 
\begin{eqnarray}
&&F = \frac{N|V|^2}{J} - \lambda N q_0 \\
&&\qquad + N\int_{-\infty}^\infty \frac{dz}{\pi} n_{\mathrm{F}}(z) \tan^{-1} %
\Big( \frac{-|V|^2 G^{\prime \prime }(z)}{z-\lambda - |V|^2 G^{\prime }(z)}%
\Big) ,  \notag
\end{eqnarray}
and for the stationarity conditions, Eq.~(\ref{Eq:stat}), 
\begin{eqnarray} 
\label{eq:jkfinalsingle}
&&\hspace{-1cm}\frac{1}{\jk} = \int_{-\infty}^{\infty}  \frac{dz}{\pi} \frac{\nf(z) G''(z)(z-\lambda)}
{(z-\lambda - |V|^2 G'(z))^2 + |V|^4 (G''(z))^2},
\\
&&\hspace{-1cm}q_0 = -  \int_{-\infty}^{\infty}  \frac{dz}{\pi} \frac{\nf(z) |V|^2 G''(z)}
{(z-\lambda - |V|^2 G'(z))^2 + |V|^4 (G''(z))^2},
\label{eq:stationaryconstraintapp}
\end{eqnarray}
which can be evaluated numerically to determine $V$ and $\lambda$ as a
function of $T$ and $\Delta$.

The Kondo temperature $\tk$ is defined by the temperature at which
$V^2 \to 0$ continuously; at such a point, the constraint Eq.~(\ref{eq:stationaryconstraintapp})
requires $\lambda \to 0$.  Here, we are interested in finding the pairing $\Delta$ at which
$\tk \to 0$; thus, this is obtained by setting $T = V = \lambda = 0$ in 
Eq.~(\ref{eq:jkfinalsingle}): 
\begin{eqnarray}
\frac{1}{J} &=& \int_{-D-\mu}^0 \frac{dz}{\pi} \frac{-\pi \rho_d(z)}{z}, \\
&=& - \rho_0 \log \frac{\Delta}{D+\mu} + \rho_0,  \label{eq:criticaljsingle}
\end{eqnarray}
where, for simplicity, in the final line we approximated $\rho_d(z)$ to be
given by 

\bea
\rho_d(\omega) \simeq\begin{cases}    
 \rho_0 |\omega|/\Delta,
& \text{for $|\omega|< \Delta$,}\cr
\rho_0  ,
& \text{for $|\omega|> \Delta$},\cr
\end{cases}
\label{rhodapprox}
\eea
that captures the essential features (except for the narrow peak near $\omega = \Delta$) 
of the true DOS of a d-wave superconductor, with
$\rho_0$ the (assumed constant) DOS of the underlying conduction band.

The solution to Eq.~(\ref{eq:criticaljsingle}) is: 
\begin{equation}
\Delta_{*}=(D+\mu )\exp \big[1-\frac{1}{\rho _{0}J}\big],
\end{equation}
showing a destruction of the Kondo effect for 
$\Delta \rightarrow \Delta_{*} $, as $V\rightarrow 0$ continuously, thus separating
the Kondo-screened (for $\Delta<\Delta_*$) from the local moment (for $\Delta>\Delta_*$)
phases.


\begin{thebibliography}{99}
\bibitem{Hewson} See, e.g., A.C. Hewson, \textit{The Kondo Problem to Heavy
Fermions\/} (Cambridge University Press, Cambridge, England, 1993). 

\bibitem{Coleman01} P. Coleman, C. Pepin, Q. Si and R. Ramazashvili, 
Journ. of Phys: Cond. Mat. \textbf{13}, R723 (2001). 

\bibitem{Nozieres98} P. Nozi\`{e}res, Eur. Phys. J. B \textbf{6}, 447
(1998). 
\bibitem{pseudogap}
By \lq\lq pseudogap\rq\rq, we are of course referring to the nodal structure of $d$-wave 
pairing, not the pseudogap regime of the high-$\tc$ cuprates. 
\bibitem{Brugger93} T. Brugger, T. Schreiner, G. Roth, P. Adelmann, and G. Czjzek, 
Phys. Rev. Lett. \textbf{71}, 2481 (1993). 
\bibitem{Withoff90} D. Withoff and E. Fradkin, Phys. Rev. Lett. \textbf{64},
1835 (1990). 
\bibitem{Borkowski92} L.S. Borkowski and P.J. Hirschfeld, Phys. Rev. B 
\textbf{46}, 9274 (1992). 
\bibitem{Ingersent96} K. Ingersent, Phys. Rev. B \textbf{54}, 11936 (1996).
\bibitem{Ingersent98} C. Gonzalez-Buxton and K. Ingersent, Phys. Rev. B 
\textbf{57}, 14254 (1998).
\bibitem{Fritz04} M. Vojta and L. Fritz, Phys. Rev. B \textbf{70}, 094502
(2004).
\bibitem{Fritz} L. Fritz and M. Vojta, Phys. Rev. B \textbf{70}, 214427
(2004). 
\bibitem{Vojtareview} M. Vojta, Philos. Mag. \textbf{86}, 1807 (2006).
\bibitem{Anderson} P.W. Anderson, J. Phys. C \textbf{3}, 2439 (1970). %
\bibitem{Tsuei00} C.C. Tsuei and J.R. Kirtley, Phys. Rev. Lett. \textbf{85},
182 (2000). 
\bibitem{Prozorov00} R. Prozorov, R.W. Giannetta, P. Fournier, R.L. Greene, 
Phys. Rev. Lett. \textbf{85}, 3700 (2000). 
\bibitem{Hien98} N.T. Hien,
V.H.M. Duijn, J.H.P. Colpa, J.J.M. Franse, and A.A. Menovsky, 
Phys. Rev. B \textbf{57}, 5906 (1998). 
\bibitem{Fulde93} P. Fulde, V. Zevin, and G. Zwicknagl, Z. Phys. B \textbf{92%
}, 133 (1993). 
\bibitem{Khaliullin95} G. Khaliullin and P. Fulde, Phys. Rev. B \textbf{52},
9514 (1995).
\bibitem{Hofstetter00} W. Hofstetter, R. Bulla, and D. Vollhardt, Phys. Rev.
Rev. Lett. \textbf{84}, 4417 (2000).
\bibitem{Huang90} Q. Huang, J.F. Zasadzinsky, N. Tralshawala, K.E. Gray, D.G. Hinks, J.L. Peng and R.L. Greene, 
Nature \textbf{347}, 369 (1990). %
\bibitem{Henggeler98} W. Henggeler and A.Furrer, Journ. of Phys. Cond. Mat. 
\textbf{10}, 2579 (1998).
\bibitem{Bala98} J. Ba\l a, Phys. Rev. B \textbf{57}, 14235  (1998).
\bibitem{Petrovic} C. Petrovic, P.G. Pagliuso, M.F. Hundley, R. Movshovich, J.L. Sarrao, 
J.D. Thompson, Z. Fisk and P. Monthoux, J. Phys. Condens. Matter 
\textbf{13}, L337 (2001). 
\bibitem{largeN} See, e.g., D.M. Newns and N. Read, Adv. Phys. \textbf{36},
799 (1987); P. Coleman, Phys. Rev. B \textbf{29}, 3035 (1984); A. Auerbach
and K. Levin, Phys. Rev. Lett. \textbf{57}, 877 (1986); A.J. Millis and P.A.
Lee, Phys. Rev. B \textbf{35}, 3394 (1987). 
\bibitem{Shiba} H. Shiba and P. Fazekas, Prog. Theor. Phys. Suppl. \textbf{%
101}, 403 (1990). %
\bibitem{Optical} S. V. Dordevic, D. N. Basov, N. R. Dilley, E. D. Bauer, and M. B. Maple,
Phys. Rev. Lett. \textbf{86}, 684 (2001); L. Degiorgi, F.B.B. Anders, and G. Gr\"uner,
Eur. Phys. J. B \textbf{19}, 167 (2001); J.N. Hancock. T. McKnew, Z. Schlesinger, J.L. Sarrao, and
Z. Fisk, 
\prl {\bf 92}, 186405 (2004). 
\bibitem{Coleman} P. Coleman, in \textit{Lectures on the Physics of Highly
Correlated Electron Systems VI\/}, F. Mancini, Ed., American Institute of
Physics, New York (2002), p. 79 - 160. 
\bibitem{Yamada} K. Yamada and K. Yosida, Prog. Theor. Phys. \textbf{76},
621 (1986). 
\bibitem{Fertig77} W. A. Fertig, D. C. Johnston, L. E. DeLong, R. W.
McCallum, M. B. Maple, and B. T. Matthias, Phys. Rev. Lett. \textbf{38}, 987
(1977).
\end{thebibliography}
\end{document}